\documentstyle[12pt]{article}
\begin{document}
\baselineskip 10mm

\centerline{\large {\bf The effect of interstitial clusters}}
\centerline{\large {\bf
and vacancies on the STM image of graphite}}

\vskip 4mm

\centerline{Arkady V. Krasheninnikov and Vladimir F. Elesin}

\vskip 2mm

\centerline{\it Moscow State Engineering Physics Institute
(Technical University),}
\centerline{\it Moscow 115409, Russia}

\vskip 2mm

\thispagestyle{empty}

\centerline{\bf Abstract}
\begin{quotation}
Making use of the tight-binding Green's function technique,
we have calculated the STM images of graphite with surface and sub-surface
defects, while taking into account the relaxation of the lattice due to
defects. We have demonstrated that two different
physical mechanisms may result in the formation of hillocks in the STM images:
buckling of the graphite surface due to interstitials between the uppermost
graphite layers and the enhancement of the electron density of states close to
the Fermi energy on the carbon atoms in the vicinity of vacancies. Our results
indicate that small hillocks may originate both from the interstitial clusters
and from the vacancies. By contrast, however, large hillocks in excess of 10
\AA~ in diameter can be caused only by interstitial clusters.  \end{quotation}

{\bf Keywords}: Computer simulations, Graphite, Scanning tunneling microscopy;
Surface defects

{\bf Corresponding author}: Dr. Arkady V.Krasheninnikov, Moscow State
Engineering
Physics Institute (Technical University), Kashirskoe shosse 31,
Moscow 115409, RUSSIA.

Phone: (095) 324 10 81; FAX: (095) 324 21 11;

E-mail: arc@supercon.mephi.ru; arckrash@glasnet.ru

\newpage
Scanning tunneling microscopy (STM) is a powerful tool for
investigating surfaces with defects inasmuch as it gives real-space images
of surfaces with atomic resolution.
Radiational defects near the surface profoundly effect STM images of
graphite.
Appearances of hillock-like features in STM images of
graphite after irradiation with noble gas
ions have been reported repeatedly
\cite{Porte,Bolse,Coratger,Hahn,Marton,Preprt}.

These hillocks (or bumps or protrusions), which reveal themselves as a
substantial increase in the tunneling current, have an average lateral diameter
of from several {\AA} up to tens of {\AA}, with the
typical height of hillocks constituting 1-5 {\AA}.
Since these structures have never been observed on
graphite surfaces free of defects, they are associated
with defects created by incident ions near the surface.

Despite numerous studies, however, the actual nature of hillock-like features
remains controversial.  Hillocks may result from local buckling of the
surface due to internal stresses induced by subsurface defects
\cite{Porte} and- -more specifically- -by carbon interstitial clusters
between the graphitic layers \cite{Bolse,Nordlund}, i.e.  within the framework
of this assumption the hillocks can be interpreted topographically.

Recent molecular dynamics (MD) simulations \cite{Nordlund,Openov} provide
evidence in favor of this interpretation. It has been
demonstrated that ion irradiation of graphite may result in the formation of
interstitial clusters under the graphite surface and in the
buckling of the surface layer.

The physical mechanism for interstitial cluster formation is proposed in
\cite{Openov}.
The interstitials created by ion impact lie predominantly
between graphite layers  \cite{Nordlund}
and have a high mobility
parallel to the layers. At the same time, as has been shown in \cite{Openov} by
the MD method, an interstitial stretches the graphite lattice within its
vicinity, with other interstitials in that same interlayer region
tending to be attracted to the stretched region.  Thus,
due to the minimization of the total buckling of the graphite layers,
the formation of
interstitial clusters lowers the deformation energy, as compared to the case of
interstitials being spatially separated.

A different interpretation \cite{Hahn} of
hillock-shaped features
is that the surface atomic vacancy is seen as a protrusion in
STM images of graphite because of the vacancy-induced enhancement
of an electron density corresponding to the states close to the
Fermi energy ($E_F$) on the carbon atoms near the vacancy.
Inasmuch as
STM probes only some of the electronic states of the sample within a
certain energy range near $E_F$ determined by the bias voltage $V_{bias}$,
the
enhancement of electron density leads to an increase in the tunneling current.
Real geometry of the surface remains practically flat.

It should also be noted that the hillocks may be explained by
incident ions trapped near the graphite surface
\cite{Marton}, but, in many cases (e.g., in the case of incident ions
with high energy), the formation of hillocks seems to
be associated with carbon atoms only.  Adsorbate effects are believed
\cite{Hahn} to be either absent or relatively negligible. Therefore, we
restrict our consideration only to cases of carbon interstitials and vacancies.

The lack of a clear understanding of the nature of the hillocks may be
due in part to the comparatively small number of theoretical works on
the subject.  The effect of lattice vacancies on the STM image of graphite
was studied in \cite{TBVAC} by the tight-binding (TB) 
molecular orbital method and in \cite{Takeuchi} using the first-principles
plane wave formalism.
The results of
simulations support the interpretation that vacancies may lead to the formation
of hillocks.

However, we are not aware of any
theoretical works dealing with the simulation of the STM images of graphite
in the presence of interstitials near the surface.

In this paper, we simulate the STM images of the graphite surface having
defects such as interstitials and vacancies making use of Green's function
calculations based on the TB model.
Our main goal is to clarify whether the vacancies and
interstitial clusters under the graphite surface result in the
formation of protrusions in the STM images.

In general, in our simulations we adopt the technique used in
\cite{McKinnon}.
To describe the graphite electronic structure, we employ the parametrization of
the TB Hamiltonian for carbon systems proposed in \cite{Xu}.
Since the density of states near $E_F$ in
graphite is determined by $\pi$-states perpendicular to the graphite layers
\cite{McKinnon, Tomanek}, we take into consideration only $2p_z$
electrons.
The tip was modeled as the final
atom of a semi-infinite, one-dimensional chain.
To the first order in the
tip-surface interaction, the tunneling current $I$ as a function of the tip
coordinates in the $(x,y)$ plane (parallel to the graphite surface) and
the tip height $h$ may be written as follows \cite{McKinnon}:
\begin{equation}
\label{trivial}
I(x,y,h)=\frac{2\pi e}{\hbar}\int_{E_F-eV_{bias}}^{E_F}\sum_{i}|V_i(x,y,h,)|
^2 \rho_{tip}(E) \rho_{surf}(r_i,E)dE,
\end{equation}
where $V$ is the tip-surface matrix element, the sum over $i$ runs
over graphite
surface sites effected by the tip, $\rho_{surf}(E)$ and $\rho_{tip}(E)$
are the  local densities of states (LDOS) of the non-interacting surface and
tip, respectively.

To account for the deformation of the graphite lattice near the
defects, the MD method was employed. Details of calculations are given in
\cite{Openov}.  Graphite was modeled by the crystallite consisting of six
graphite layers, each composed of 1096 atoms.
Crystallites
with different numbers of atoms were also used, but the results proved
to be practically independent of the crystallite size.
Interstitials were
initially distributed randomly over the crystallite between the uppermost
layers.  Ten initial variants were considered for each number of interstitials.
The tunneling current was calculated after the relaxation of the lattice and
interstitial cluster formation due to deformation interaction.

We have calculated the STM images of the graphite surface near both a single
interstitial as well as clusters composed of 2--10 interstitials.
Fig.1 (a) shows an isometric plot of the variation in the tip height $h$
for a scan across ($x,y$) near ten-interstitial cluster on the graphite surface
at a current of 4 nA and $V_{bias}=0.08$ V. Fig.1(b) depicts a contour map
corresponding to Fig.1(a) and simulating the STM image.
A dramatic protrusion above the interstitial cluster is
evident.  This hillock is due to a graphite surface buckling($\approx
1 $ \AA) which results in a corresponding growth of $V$ and in an increase in
the tunneling current. The graphite LDOS does not change substantially near the
cluster.  The positions of maxima and minima of the tip height near the cluster
correspond to those in the defect-free case (in graphite every second atom is
observable due to electronic effects \cite{Tomanek}). Slight anisotropy in the
corrugation ($\approx 0.05$ \AA~) near the cluster is associated with the
non-uniform distribution of electron density because of the anisotropic form of
the interstitial cluster, see Fig.1(b).

Hillock-like features were observed for all clusters, but the height and
lateral diameter of protrusions were dependent on the number of interstitials
in the cluster $N_{int}$ as well as on cluster shape. The average
diameter of hillocks $\bar d$ as a function of $N_{int}$ is shown in Fig.2. It
is seen that  $\bar d$ increases linearly with $N_{int}$. The average height
also grows with $N_{int}$, but rather slowly ($\bar h \approx 1$ \AA~ for
$N_{int}=1$ and $\bar h \approx 1.5$ \AA~ for $N_{int}=10$), which may be
associated with
the two-dimensional character of interstitial clusters.

Let us proceed to the effect of vacancies on the STM image.
Fig.3 shows the variation in the tip
height $h$ for a scan on the graphite surface near a single
vacancy positioned on the $\beta$ surface site
(there is no atom below the $\beta$ site in the second graphite layer).
The tunneling current and $V_{bias}$
were the same as in the case of interstitials.  As follows from Fig.3 and
although real geometry remains flat, the vacancy results in the formation
of a hillock with the height of $\approx 0.8$ \AA.
The increase in $h$ is governed by
the enhancement of the LDOS at $E_F$ on the atoms
in the immediate vicinity of the
vacancy, which is in agreement with the results of \cite{TBVAC}.

The superstructure (SS) $\sqrt{3}\times\sqrt{3}R30^\circ$ is seen in
the image. The SS was also evident in our calculated images near single
interstitials.
Similar SS were experimentally imaged next to some hillocks in
\cite{Porte}. Our calculations confirm the interpretation
\cite{Interfer} that
these SS are actually a purely electronic effect due to point defects
 and are not indicative of any surface reconstruction.
Note that SS are absent in Fig.1 since interstitial clusters result in a
long-range perturbation of the lattice.

Qualitatively similar results were obtained for the vacancies in the $\alpha$
positions, divacancies, and trivacancies.
The lateral size of the vacancy-induced hillock depends slightly on the number
of vacancies in the system, see Fig.2. Therefore, our results indicate
that large, experimentally-observed hillocks exceeding 10 \AA~ in diameter
can be caused only by the interstitial clusters.

In summary, we have calculated  the STM images of graphite having surface and
sub-surface defects. Two different physical
mechanisms may give rise to the appearance of hillocks in the STM
images: buckling of the graphite surface due to interstitials between the
uppermost graphite sheets and the enhancement of the partial charge density of
states on the carbon atoms near vacancies. Small hillocks ($< 10$ \AA~  in
diameter) may originate from both the interstitial clusters and
vacancies. However, large hillocks ($> 10$ \AA) may be caused only by
interstitial clusters. Thus, our results confirm the mechanism
\cite{Openov} of interstitial cluster formation in the irradiated graphite due
to deformation interaction.

We would like to thank Dr. L.A. Openov and Prof. L.A. Suvorov for their helpful
comments. A.V.K. is indebted to V.M\"uller for discussion.
 This work has been supported by the International Science and Technology
Center (Project 467) and by the Russian Federal Program "Integration" (Project
No AO133).

\newpage
{\bf Figure captions}

Fig.1. Constant current mode of operation of STM:
(a) Variation in tip height $h$ for a scan across $x,y$ on the graphite surface
near a ten-interstitial cluster at a current of 4nA and $V_{bias}=0.08$ V.
(b) The contour map of 1(a). Dots correspond to positions of carbon atoms
in the surface layer, squares correspond to carbon atoms in the interstitial
cluster.

Fig.2. The average diameter $\bar d$ of the hillock as a function of the number
of interstitials and vacancies in the system.

Fig.3. The same as in Fig.1, but for a single vacancy.

\end{document}